\documentclass[conference]{IEEEtran}
\IEEEoverridecommandlockouts
\usepackage{cite}
\usepackage{amsmath,amssymb,amsfonts}
\usepackage{algorithmic}
\usepackage{graphicx}
\usepackage{textcomp}
\usepackage{xcolor}

\usepackage{algorithm}
\usepackage{url}
\usepackage{hyperref}
\usepackage[caption=false,font=normalsize,labelfont=sf,textfont=sf]{subfig}
\usepackage{soul}
\usepackage{multirow}
\usepackage{svg}
\usepackage{balance}
\usepackage{comment}

\makeatletter
\newcommand{\linebreakand}{%
  \end{@IEEEauthorhalign}%
  \hfill\mbox{}\par%
  \mbox{}\hfill\begin{@IEEEauthorhalign}}
\makeatother

\def\BibTeX{{\rm B\kern-.05em{\sc i\kern-.025em b}\kern-.08em
    T\kern-.1667em\lower.7ex\hbox{E}\kern-.125emX}}
\begin{document}

\title{Investigation of Time-Frequency Feature Combinations with Histogram Layer Time Delay Neural Networks
\thanks{DISTRIBUTION STATEMENT A. Approved for public release. Distribution is unlimited. This material is based upon work supported by the Under Secretary of Defense for Research and Engineering under Air Force Contract No. FA8702-15-D-0001. Any opinions, findings, conclusions or recommendations expressed in this material are those of the author(s) and do not necessarily reflect the views of the Under Secretary of Defense for Research and Engineering. Code is publicly available at this GitHub \url{https://github.com/Advanced-Vision-and-Learning-Lab/HLTDNN_Feature_Combination}.}
}

\author{
  \IEEEauthorblockN{\hspace*{-10mm} 1\textsuperscript{st} Amirmohammad Mohammadi}%
  \IEEEauthorblockA{\hspace*{-10mm}
    \textit{Department of Electrical and Computer Engineering}\\
    \hspace*{-10mm}\textit{Texas A\&M University}\\
    \hspace*{-10mm}College Station, TX, USA\\
    \hspace*{-10mm}amir.m@tamu.edu}%
  \and
  \IEEEauthorblockN{\hspace*{6mm}2\textsuperscript{nd} Iren\'e Masabarakiza}%
  \IEEEauthorblockA{\hspace*{6mm}
    \textit{Department of Electrical and Computer Engineering}\\
    \hspace*{6mm}\textit{Texas A\&M University}\\
    \hspace*{6mm}College Station, TX, USA\\
    \hspace*{6mm}imasabo2k18@tamu.edu}%
  \linebreakand
  \IEEEauthorblockN{\hspace*{-8mm} 3\textsuperscript{rd} Ethan Barnes}%
  \IEEEauthorblockA{\hspace*{-8mm}
    \textit{Department of Electrical and Computer Engineering}\\
    \hspace*{-8mm}\textit{Texas A\&M University}\\
    \hspace*{-8mm}College Station, TX, USA\\
    \hspace*{-8mm}ecbarnes@tamu.edu}%
  \and
  \IEEEauthorblockN{\hspace*{6mm}4\textsuperscript{th} Davelle Carreiro}%
  \IEEEauthorblockA{%
  \hspace*{6mm}%
    \textit{Massachusetts Institute of Technology Lincoln Laboratory}\\
    \hspace*{6mm} Lexington, MA, USA\\
    \hspace*{6mm} davelle.carreiro@ll.mit.edu}%
  \linebreakand
  \IEEEauthorblockN{\hspace*{-15mm}5\textsuperscript{th} Alexandra Van Dine}%
  \IEEEauthorblockA{\hspace*{-15mm}
    \textit{Massachusetts Institute of Technology Lincoln Laboratory}\\
    \hspace*{-15mm} Lexington, MA, USA\\
    \hspace*{-15mm} alexandra.vandine@ll.mit.edu}%
  \and
  \IEEEauthorblockN{\hspace*{1mm} 6\textsuperscript{th} Joshua Peeples}%
  \IEEEauthorblockA{\hspace*{2mm} 
    \textit{Department of Electrical and Computer Engineering}\\
    \hspace*{2mm} \textit{Texas A\&M University}\\
    \hspace*{2mm} College Station, TX, USA\\
    \hspace*{2mm} jpeeples@tamu.edu}%
}

\maketitle

\begin{abstract}
While deep learning has reduced the prevalence of manual feature extraction, transformation of data via feature engineering remains essential for improving model performance, particularly for underwater acoustic signals. The methods by which audio signals are converted into time-frequency representations and the subsequent handling of these spectrograms can significantly impact performance. This work demonstrates the performance impact of using different combinations of time-frequency features in a histogram layer time delay neural network. An optimal set of features is identified with results indicating that specific feature combinations outperform single data features.
\end{abstract}

\begin{IEEEkeywords}
deep learning, underwater acoustic, histogram layers, feature engineering
\end{IEEEkeywords}

\section{Introduction}
\label{sec:intro}

Driven by an ability to effectively process large volumes of data, deep learning (DL) has been successfully used for acoustic classification tasks \cite{qu2022acoustic, mu2021environmental}. One application of sound classification is Underwater Acoustic Target Recognition (UATR), where acoustic signals are used to classify underwater objects. This technology has various applications in marine environments such as biological pattern of life determination, search and rescue, seabed mapping, and shipping traffic monitoring \cite{testolin2020detecting, beckler2022multilabel}. DL methods are being widely developed due to their effective ability in processing time-frequency features generated from raw input signals and performing end-to-end tasks such as classification for underwater acoustic signals \cite{feng2022transformer, tian2021deep, wang2023underwater}. While DL can automatically extract and learn features, the transformation and manipulation of audio signals is an important step of feature engineering \cite{Purwins2019}.

Generally, time-domain sound signals are first converted into time-frequency representations, known as spectrograms, before being processed by artificial neural networks \cite{Zaman2023}. This is useful because spectral representations contain unique acoustic signatures of underwater objects, allowing models like convolutional neural networks (CNNs) to learn from these patterns. Success with this method has been demonstrated in image analysis tasks \cite{liu2021underwater}.

Images can be decomposed into statistical and structural texture features \cite{ramola2020study, OZSEVEN201870, ji2022structural}. Spectrograms can also be represented through these two texture types. Statistical texture features provide information about the texture \cite{trevorrow2021examination}, using statistical measures that usually capture the distribution. Structural texture features, on the other hand, are the spatial patterns of textures \cite{peeples2021histogram}. Models like and time delay neural networks (TDNNs) are very effective at capturing spatial and temporal arrangements, or structural textures in imagery such as spectrograms \cite{ji2022structural}. However, these models may not fully take into account the statistical features within spectrograms. 

\begin{figure*}[htb]
\centering
\includegraphics[width=0.85\textwidth]{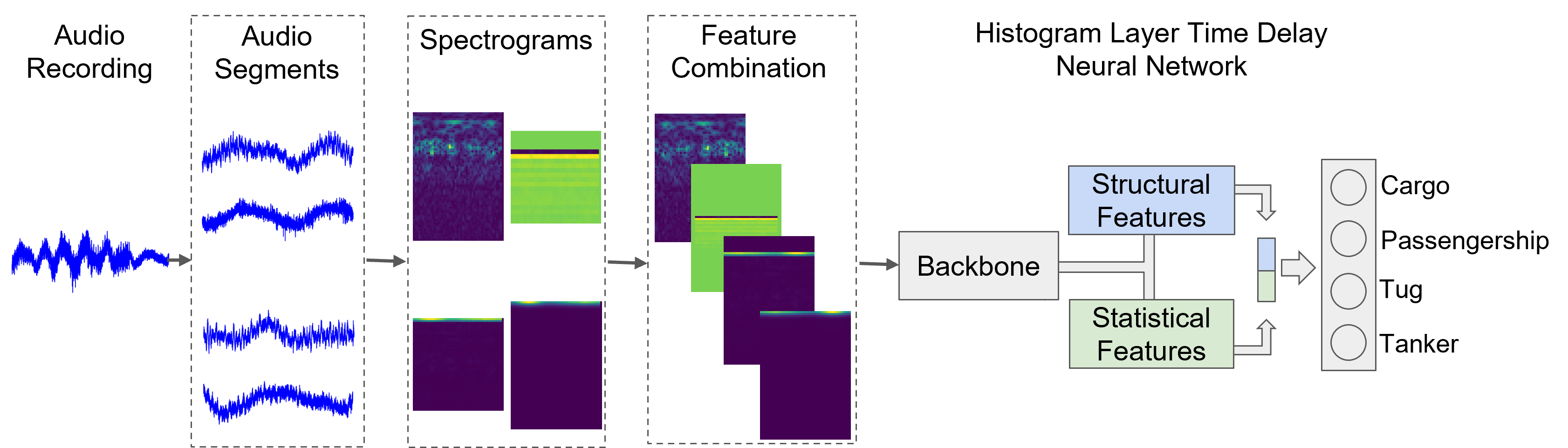}
\caption{Overall workflow. The audio segments are first partitioned into individual segments. Each audio segment is then used to compute an individual feature of varying time and frequency resolutions. We introduce an adaptive padding layer to make each spectrogram the same size. The features are then concatenated along the channel dimension and processed by the HLTDNN model for classification.}
\label{fig:flow}
\end{figure*}

Histogram layers are designed to model the distribution of feature values within local regions of images \cite{peeples2021histogram}. The motivation is that histogram layers directly model statistical features by characterizing spatial distributions of features as opposed to other deep learning approaches such as CNNs. The integration of these layers into TDNNs, referred to as histogram layer time delay neural network (HLTDNN), \cite{ritu2023histogram} improved classification performance for UATR. This addition improved capture of statistical context, thereby providing a more accurate representation of the feature distributions across local spatial regions \cite{peeples2021histogram}. Prior work \cite{ritu2023histogram} on the HLTDNN used single spectrogram data inputs for classification. 

The representation of audio signals is a difficult problem yet crucial to the success of machine learning algorithms \cite{sharma2020trends}. Related work \cite{arias2021multi} has demonstrated that combining three types of spectrograms, mel-frequency spectrograms, gammatone-frequency spectrograms (Cochleagrams), and continuous wavelet transforms (CWT), can improve the detection of speech deficits in cochlear implant users. The task of acoustic feature engineering, particularly for underwater environments, still remains challenging \cite{aslam2024underwater}. While a prior study showed feature fusion to be effective in improving UATR performance \cite{ke2020integrated}, it focused on a smaller UATR dataset, ShipsEar \cite{santos2016shipsear}. A related effort \cite{zhang2022multi} used a larger dataset, DeepShip \cite{irfan2021deepship}, to show that feature fusion could improve performance, but did not investigate the impact of feature selection for three spectrogram features used. Additionally, spectrograms for the inputs are restricted to be the same size and cannot use different hyperparameters (\textit{e.g.}, window length, hop length) for each spectrogram which may unintentionally restrict the unique information captured by each feature.

We leverage the HLTDNN in our work, and focus on extending the application of spectrogram techniques by investigating various combinations of acoustic data features to find an optimal set of features which yields improved classification performance. This approach is driven by the hypothesis that a combination of acoustic data features could provide a richer representation of the input signals \cite{Chi2019}. To our knowledge, this is the first investigation of feature combinations for the DeepShip dataset using HLTDNNs.

\section{Method}
\label{sec:meth}

\subsection{Review of the HLTDNN architecture and the histogram layer operation}
\label{ssec:subheadmeth1}

The HLTDNN's \cite{ritu2023histogram} backbone consists of multiple convolution layers that have the rectified linear unit (ReLU) activation function, followed by max pooling layers. The histogram layer is positioned in parallel to another convolutional block. The input into both the histogram layer and convolutional blocks are from the the fourth convolutional block. The histogram layer captures statistical texture information while the convolutional block focuses on structural texture information. The statistical and structural features are concatenated together and passed into the output classification layer \cite{peeples2021histogram,ritu2023histogram}.

Let $\mathbf{X} \in \mathbb{R}^{M \times N \times D}$ be the input to a histogram layer, where $M, N$ are spatial dimensions and $D$ is feature dimensionality. Applying a histogram layer with $B$ bins yields output $\mathbf{Y} \in \mathbb{R}^{R \times C \times B \times D}$, where $R, C$ are output spatial dimensions determined by kernel size $S \times T$:

\begin{equation}
    Y_{rcbd} = \frac{1}{ST} \sum_{s=1}^{S} \sum_{t=1}^{T} e^{-\gamma_{bd}^2(x_{r+s,c+t,d} - \mu_{bd})^2}
\end{equation}

\noindent Here, $\mu_{bd}$ and $\gamma_{bd}$ are learnable bin centers and widths. The layer processes each input feature dimension, producing $BD$ histogram feature maps. It uses radial basis functions to determine feature contributions to bins, capturing feature value distributions and adding statistical context to the model.

\subsection{Feature combination implementation}
\label{ssec:subhead3meth}

We propose merging different spectrogram features to potentially improve the representational power of the input data passed into the model. Given a total of $\mathcal{M}$ unique time-frequency features, the total number of unique combinations, $\mathcal{N}$, can be computed as $\mathcal{N} = 2^\mathcal{M} - 1$.
For our work, six different time-frequency features are combined to generate 63 distinct combinations (\(2^6 - 1\)). To prepare these combinations of features for input into the model, an adaptive zero padding layer is incorporated into the spectrograms. During adaptive padding, since features do not share the same height and width dimensions, they are resized to a common dimension, as to be combined without losing their time-frequency resolution by calculating the required padding for each side of a feature map to match the largest height and width among all features. Padding is applied symmetrically to maintain the feature's original structure and position within its new dimensions. Once all features are resized to a uniform dimension, they are combined by concatenating the feature maps along the channel dimension. The overall workflow is shown in Fig. \ref{fig:flow}.

\section{Experimental Setup}
\label{sec:expmajhead}

In this work, the DeepShip sonar dataset is used \cite{irfan2021deepship}. This dataset has a collection of real-world sonar recordings with 265 different ships across four classes (tanker, tug, passengership, and cargo), compiled from 47 hours and 4 minutes of recordings. Recordings are captured under various sea states and noise levels. Audio signals are uniformly sampled to a frequency of 16 kHz and then segmented into intervals of three seconds. Next, six time-frequency based features are extracted: mel-frequency spectrogram (MS), mel-frequency Cepstral coefficients (MFCCs), short-time Fourier transform (STFT), gammatone-frequency Cepstral coefficients (GFCC), constant-Q transform (CQT), and variable-Q transform (VQT). Studies have demonstrated the effectiveness of these features in classifying underwater acoustic signals \cite{irfan2021deepship, guo2022underwater,ritu2023histogram}. For the extraction process, the window and hop length are set at 250 ms and 64 ms, respectively. Specifically, MS uses 44 Mel filter banks. In the case of MFCCs, there are 16 mel-frequency cepstral coefficients. Meanwhile, STFT has 48 frequency bins. For the GFCC, CQT, and VQT features, there are 64 frequency bins.

The dataset was partitioned into training, validation, and test sets with respective ratios of 70\%, 15\%, and 15\% based on \cite{ritu2023histogram}. A total of 56,468 segments were generated, distributed as 38,523 for training, 9,065 for validation, and 8,880 for testing. To prevent data leakage, segments derived from a single recording were exclusively allocated to the same partition. This ensures if a recording is chosen for training, all its segments are also chosen solely for training purposes. The HLTDNN model was evaluated with each combination of features across three runs of random initialization. The configuration for the experimental parameters was as follows: Adagrad optimizer with a learning rate of 1e-3 and a batch size of 128. A dropout rate of 0.5 was applied preceding the output classification layer to mitigate overfitting. Additionally, early stopping was implemented to stop training if no improvement in validation loss is observed after 15 epochs. The maximum epoch number was set to 150. Following \cite{ritu2023histogram}, the number of bins was set to 16 for the HLTDNN model. 
\section{Results and Discussion}
\label{sec:resultsec}

\begin{table*}[htb]
\caption{Overall performance metrics including the
Matthews Correlation Coefficient (MCC) for the top five combinations and the best single feature. The results shown are based on the average value with $\pm 1\sigma$ across the three experimental runs of random initialization. For brevity, we show the top five feature combinations. The full set of results can be found in the code repository.}
\centering
\label{tab-compare}
\begin{tabular}{|c|c|c|c|c|c|c|}
\hline
Features & Accuracy (\%) & Precision (\%) & Recall (\%) & F1 score (\%) & MCC \\ \hline
VQT, MFCC, STFT, GFCC & \textbf{66.17 $\pm$ 1.10} & \textbf{67.17 $\pm$ 1.53} & \textbf{66.17 $\pm$ 1.10} & \textbf{65.95 $\pm$ 1.17} & \textbf{0.548 $\pm$ 0.015} \\ \hline
MS, CQT, MFCC, STFT & 65.69 $\pm$ 1.68 & 66.01 $\pm$ 2.69 & 65.69 $\pm$ 1.68 & 65.00 $\pm$ 2.04 & 0.544 $\pm$ 0.023 \\ \hline
MFCC, STFT, GFCC & 65.55 $\pm$ 1.02 & 66.32 $\pm$ 1.92  & 65.55 $\pm$ 1.02 & 64.85 $\pm$ 1.16 & 0.543 $\pm$ 0.016 \\ \hline
CQT, MFCC, STFT & 65.49 $\pm$ 0.65 & 66.47 $\pm$ 0.62 & 65.49 $\pm$ 0.65 & 64.78 $\pm$ 1.07 & 0.547 $\pm$ 0.006 \\ \hline
MS, MFCC, STFT, GFCC & 64.89 $\pm$ 1.12 & 66.36 $\pm$ 1.08 & 64.89 $\pm$ 1.12 & 63.91 $\pm$ 1.19 & 0.540 $\pm$ 0.012 \\ \hline
MFCC & 59.34 $\pm$ 3.16 & 60.46 $\pm$ 1.83 & 59.34 $\pm$ 3.16 & 58.37 $\pm$ 3.88 & 0.461 $\pm$ 0.036 \\ \hline
\end{tabular}
\end{table*}

The top five feature combinations and best single feature (MFCC) performance metrics are reported in Table \ref{tab-compare}. In evaluating the performance of the model across various feature combinations, it was observed that the combination of VQT, MFCC, STFT, and GFCC features resulted in the highest classification performance across multiple metrics. This new combination achieved an accuracy of \(66.17 \pm 1.10\%\) and outperformed the highest accuracy of the single input feature case, \(59.34 \pm 3.16\%\), using MFCC alone. Therefore, an increase of \(6.83\%\) in accuracy is achieved. 
This result shows the synergistic effect of combining acoustic features for UATR.

\begin{figure}[ht]
\centering
\subfloat[]{\includegraphics[width=0.235\textwidth]{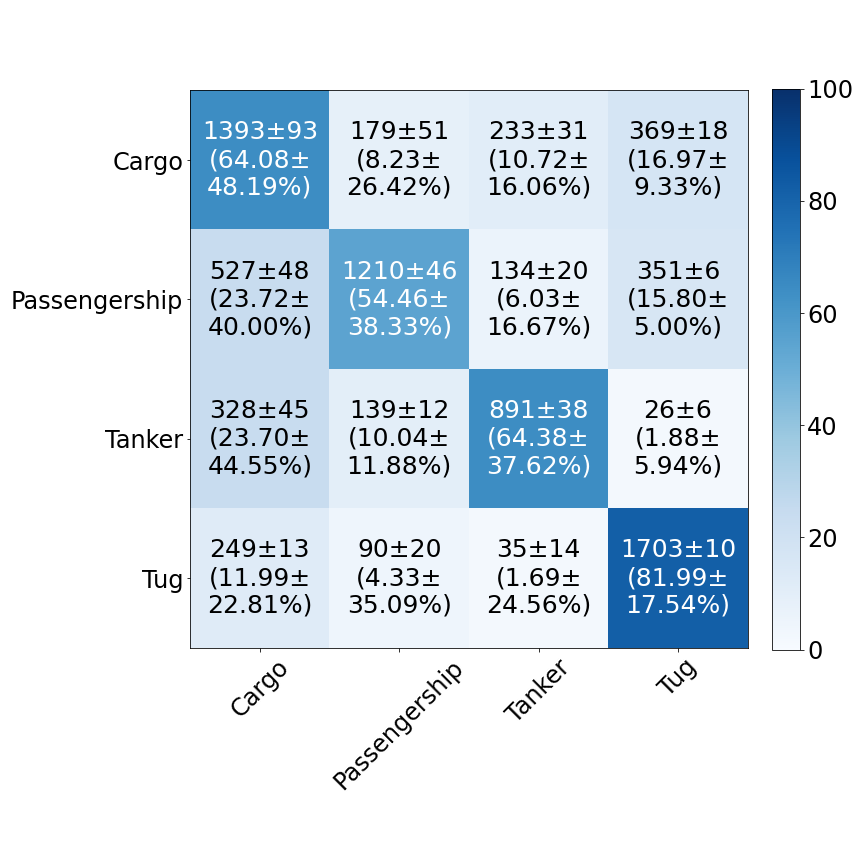}%
\label{fig:C-M1}}
\hfil
\subfloat[]{\includegraphics[width=0.235\textwidth]{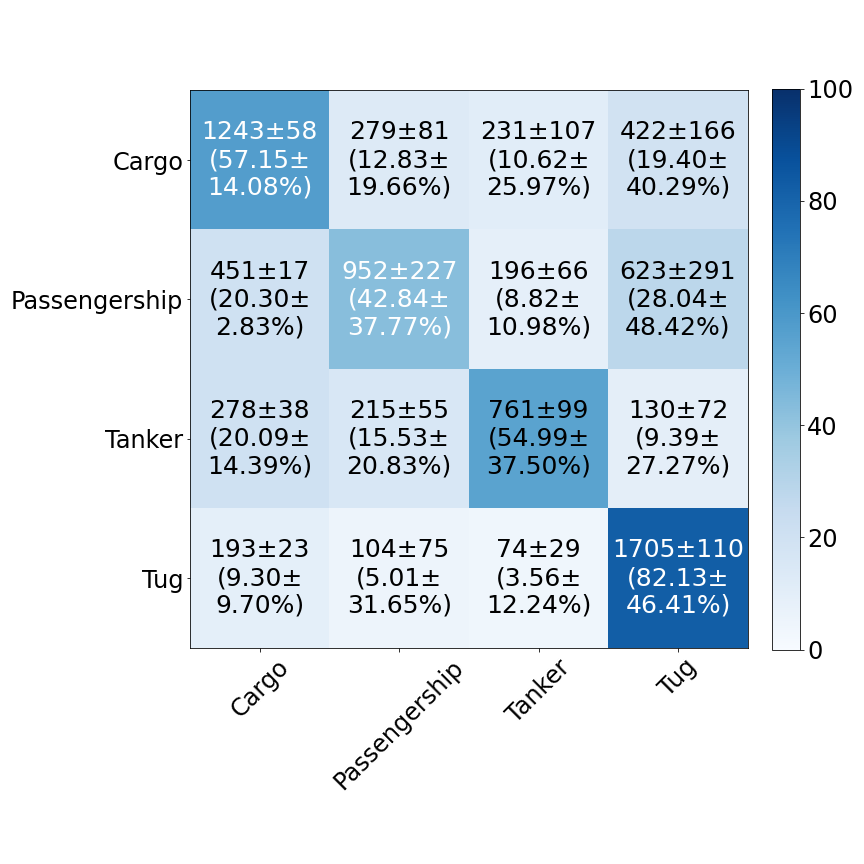}%
\label{fig:C-M2}}
\caption{Comparison of confusion matrix results for (a) best combination of features (66.17 $\pm$ 1.10\%) and (b) MFCC alone (59.34 $\pm$ 3.16\%).}
\label{fig:C-M}
\end{figure}

\begin{figure}[htb]
\centering
\subfloat[]{\includegraphics[width=0.235\textwidth]{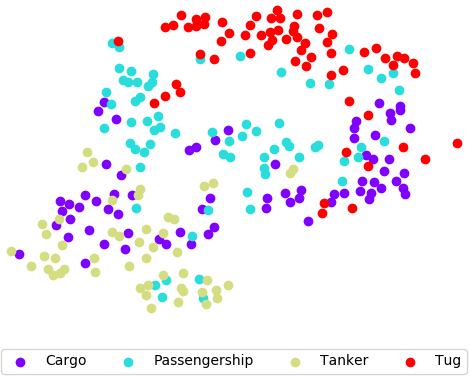}%
\label{fig:T-V1}}
\hfil
\subfloat[]{\includegraphics[width=0.235\textwidth]{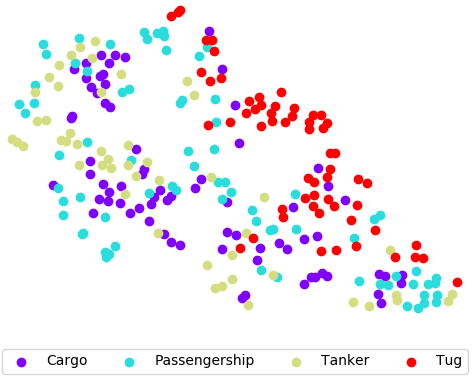}%
\label{fig:T-V2}}
\caption{t-SNE results for (a) best feature combination (FDR: 61.26 $\pm$ 4.52) and (b) MFCC (FDR: 29.04 $\pm$ 2.49). Colors represent four classes of ships. The average log Fisher Discriminant Ratio (FDR) ($\pm 1\sigma$) is also shown, with higher scores indicating more compact and better-separated classes. The best random seed is used for each feature.}
\label{fig:T-V}
\end{figure}

The best-performing feature combination was then further analyzed using a confusion matrix and t-SNE visualization to investigate classification details and feature space distribution, respectively. These visuals show the differentiation between classes. The confusion matrix and the t-SNE for both the best feature combination and the best single feature are shown in Figs. \ref{fig:C-M} and \ref{fig:T-V}, respectively. In the confusion matrices, it is noted that the combination of features led to reduced variability in model prediction. The best feature combination resulted in reduced standard deviation in correct prediction for three of the four vessel types (passengership, tanker, tug). It is also noted that the best features led to improved peformance for cargo, pssengership and tanker with comparable performance seen for the tug class when compared to MFCC alone. 

The log Fisher’s Discriminant Ratio (FDR) was computed as a measure to evaluate the separability of the features by considering the distance between class means, normalized by the variance within each class. For the best feature combination, higher FDR scores and the t-SNE visualizations in Fig. \ref{fig:T-V} show strong performance compared to the single feature where there is more overlap between classes. Specifically, passengership data points are more compact and centered for the best combination of features while tug data points are more separated. The best random seed initialization (\textit{i.e.}, experimental run with the best performance metrics) was used for each t-SNE visualization to compare the best feature combination and a single feature (MFCC). Along with fixing the random seed, the number of features from the penultimate layer is the same for the feature combination and MFCC ($\mathbb{R}^{512}$) since the HLTDNN model uses an adaptive average pooling layer in the base architecture \cite{ritu2023histogram}.

\begin{figure}[ht]
\centering
\subfloat[]{\includegraphics[width=0.155\textwidth]{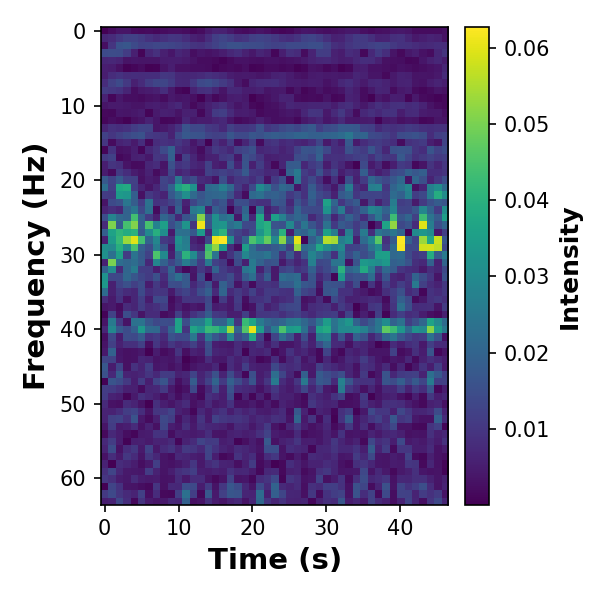}%
\label{fig:row1fig1}}
\hfil
\subfloat[]{\includegraphics[width=0.155\textwidth]{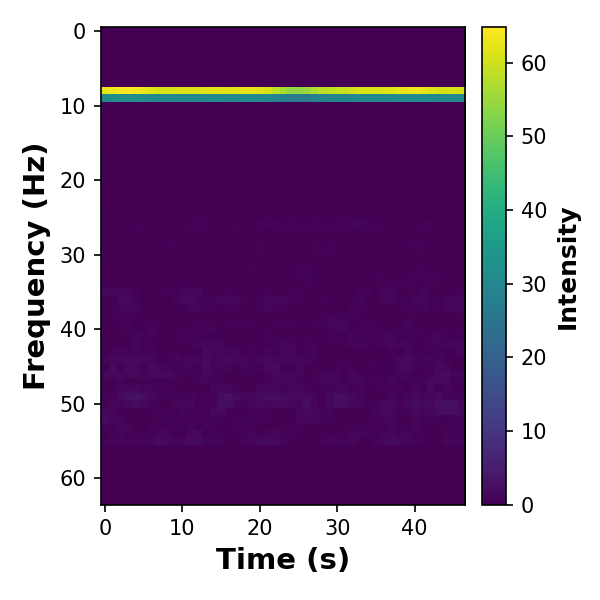}%
\label{fig:row1fig2}}
\hfil
\subfloat[]{\includegraphics[width=0.155\textwidth]{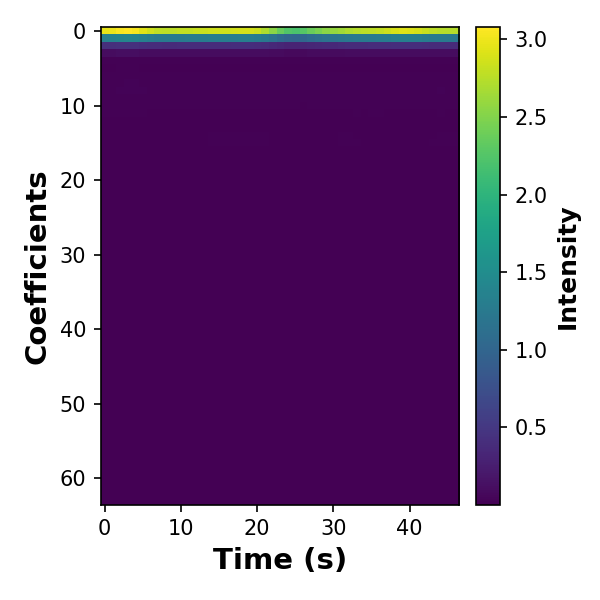}%
\label{fig:row1fig3}}
\ 
\subfloat[]{\includegraphics[width=0.155\textwidth]{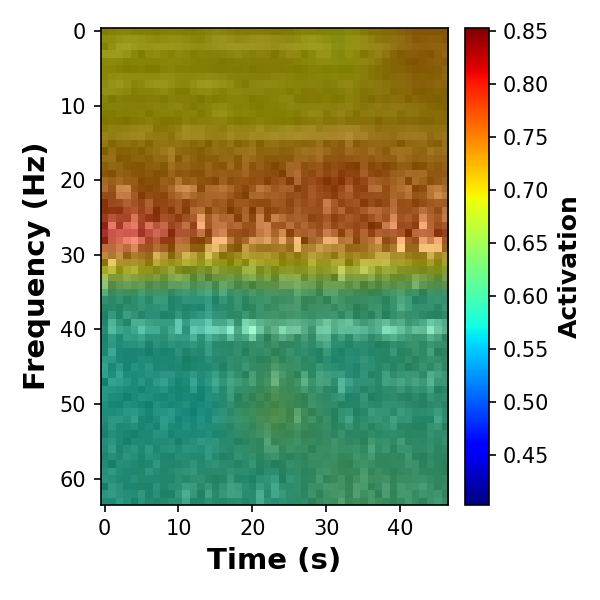}%
\label{fig:row2fig1}}
\hfil
\subfloat[]{\includegraphics[width=0.155\textwidth]{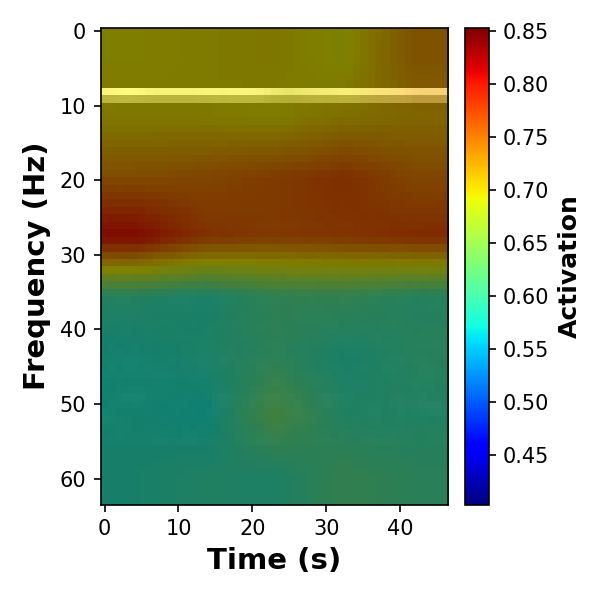}%
\label{fig:row2fig2}}
\hfil
\subfloat[]{\includegraphics[width=0.155\textwidth]{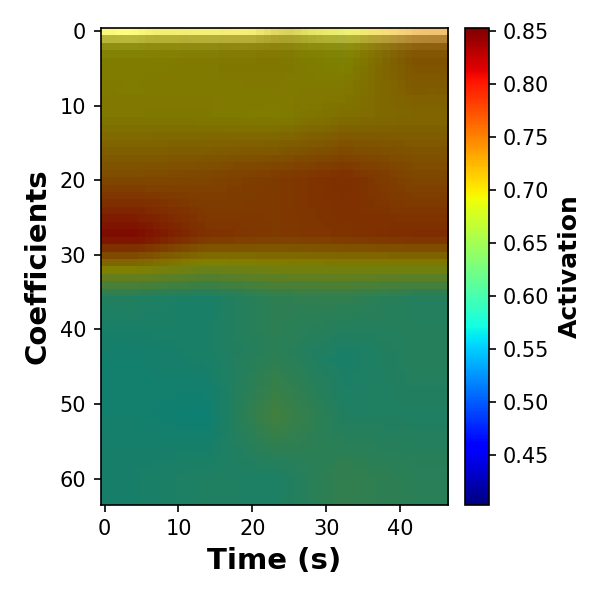}%
\label{fig:row2fig3}}
\ 
\subfloat[]{\includegraphics[width=0.155\textwidth]{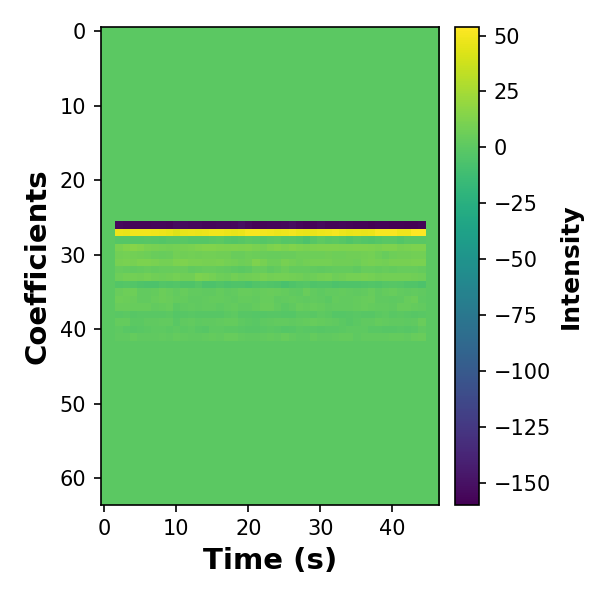}%
\label{fig:row3fig1}}
\hfil
\subfloat[]{\includegraphics[width=0.155\textwidth]{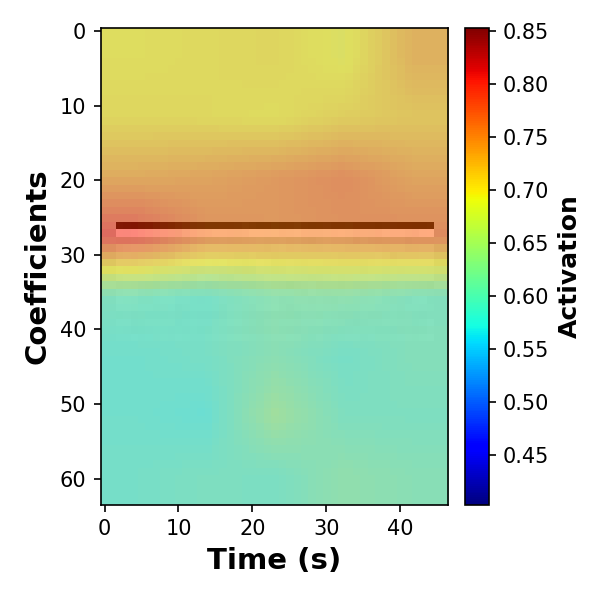}%
\label{fig:row3fig2}}
\hfil
\subfloat[]{\includegraphics[width=0.155\textwidth]{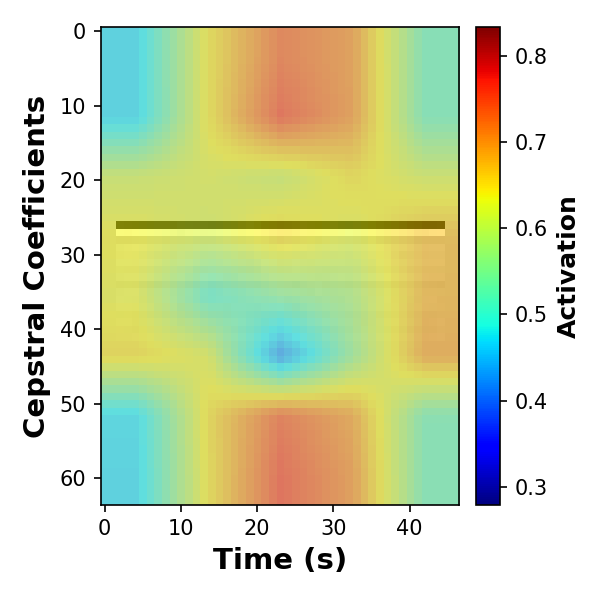}%
\label{fig:row3fig3}}
\caption{Class Activation Maps (CAM) using the best feature combination and MFCC alone for the Cargo class. The third row compares the CAM overlay on the MFCC for both models. (a) Original VQT, (b) Original STFT, (c) Original GFCC, (d) Best CAM for VQT, (e) Best CAM for STFT, (f) Best CAM for GFCC, (g) Original MFCC, (h) Best CAM, (i) MFCC CAM.}
\label{fig:composite}
\end{figure}

The reason that the model using combined features is performing better can be explained by the unique characterizations of the acoustic signal that each spectrogram contains. The repeated appearance of MFCC and STFT across the top-performing combinations shows that they are capturing more impactful signal characteristics. Each feature is sensitive to different aspects of the sound signal and for each type of ship, specific patterns can be found in the spectrograms due to the unique properties of their acoustic emissions. For example, the low-frequency content can be dominated by a propeller signature \cite{irfan2021deepship}. MFCC have proven effective in representing these sound signals, using a mel scale that approximates the human auditory system's response \cite{irfan2021deepship, Guo2022, Abdul2022}. Similarly, GFCCs are computed with a gammatone filter bank effective for the analysis of low-frequency sounds \cite{Lian2017}. VQT provides improved frequency resolution at low frequencies and better time resolution at high frequencies \cite{Sch2010, Cao2019, Sch2014}. STFT is noted for its effectiveness in analyzing narrowband signals by producing a consistent set of local spectra across time intervals \cite{jia2017rigid, Wang2023}. The reason MS and CQT are not present in the top feature combination might be due to their potential overlap in contributions with other features. Specifically, VQT somewhat covers CQT, and the information captured by MS might be sufficiently represented by the combination of MFCC, GFCC, and STFT.

Explainable artificial intelligence (XAI), specifically the FullGrad Class Activation Mapping (FullCAM) method \cite{jacobgilpytorchcam}, is used to analyze the model's decision-making process as shown in Fig. \ref{fig:composite}. For illustration, this method was applied to analyze the cargo class. By overlaying CAM on spectrograms of the best feature combinations as well as MFCC alone, the model's focus areas for classification decisions can be seen. This specific sample was correctly classified by the best feature combination model but was misclassified by the MFCC model. Colors indicate the importance a model placed on a particular component of the data feature. The CAM output difference between the best feature combination and MFCC alone shows that the model using the combined features is more focused on specific frequency bands. MFCC alone is focusing on a different range of frequencies and potentially cannot achieve the same discriminative ability. FullCAM shows how important different parts of the data are by looking at the gradients of the biases from all layers of the network, unlike other methods that focus on certain layer(s) in the network. This helps provide insight into which specific frequencies appear more important in the input spectrogram for classification as the network is focusing on these regions of the time-frequency feature throughout the model.

\section{Conclusion}
\label{sec:conc}
Our work shows the importance of feature selection to improve the performance of the HLTDNN model, particularly in the context of acoustic signal processing. The pipeline was adapted to process acoustic data feature combinations to better capture representations of the acoustic signals. It is concluded that the combination of VQT, MFCC, STFT, and GFCC outperforms other combinations especially when compared to the best single feature, MFCC. Future work includes investigating the integration of learnable feature representations, potentially through end-to-end DL models that can automatically optimize feature combinations. Additionally, while an exhaustive search of feature combinations can find the optimal solution \cite{jia2022feature}, this approach is not computationally efficient \cite{jia2022feature,peeples2018possibilistic}. Other approaches such as filter-based feature selection techniques can improve the computational cost of ``pre-screening" features instead of generating, training, and evaluating a model for each feature \cite{jia2022feature,bommert2020benchmark}.


\balance
\bibliographystyle{IEEEtran}
\bibliography{refs}


\end{document}